\documentclass[aps,prb,superscriptaddress,showpacs,citeautoscript,twocolumn]{revtex4}

\usepackage{graphicx}
\usepackage{amsmath}
\usepackage{amssymb}

\newcommand{\bnen}{\begin{equation}}
\newcommand{\eden}{\end{equation}}
\newcommand{\bean}{\begin{eqnarray}}
\newcommand{\eean}{\end{eqnarray}}
\newcommand{\bnsn}{\begin{subequations}}
\newcommand{\edsn}{\end{subequations}}
\newcommand{\f}{\frac}

\renewcommand{\vec}[1]{\text{\boldmath{$ #1 $}}}
\begin{document}
\title{Spin-dependent electron-impurity scattering in two-dimensional electron systems}

\author{A. P\'alyi}
%\email{palyi@complex.elte.hu}
\affiliation{Department of Physics of Complex Systems,
E\"otv\"os University,
H-1117 Budapest, P\'azm\'any P\'eter s\'et\'any 1/A, Hungary}

\author{J. Cserti}
%\email{palyi@complex.elte.hu}
\affiliation{Department of Physics of Complex Systems,
E\"otv\"os University,
H-1117 Budapest, P\'azm\'any P\'eter s\'et\'any 1/A, Hungary}

\date{\today}

\begin{abstract}
We present a theoretical study of elastic 
spin-dependent electron scattering caused by a charged impurity in the vicinity 
of a two-dimensional electron gas. 
We find that the symmetry properties of the spin-dependent differential scattering
cross section are different for an impurity located in the plane of the electron 
gas and for one at a finite distance from the plane.
We show that in the latter case asymmetric (`skew') scattering can arise
if the polarization of the incident electron has a finite projection on the
plane spanned by the normal vector of the two-dimensional electron gas
and the initial propagation direction. 
In specially prepared samples this scattering mechanism may give rise to a Hall-like effect in the presence of an 
\emph{in-plane} magnetic field.
\end{abstract}

 %Quantum wells 

\maketitle

%===

In quantum scattering theory one of the central quantities is the differential
scattering cross section (DSCS).
In the so-called $S$-matrix formalism it is possible to derive the
symmetry properties of the DSCS if 
the symmetries of the Hamiltonian -- the operators commuting with the
Hamiltonian -- are known\cite{ballentine,palyi-skewpaper}. 
A particular example was studied by Huang et al.\cite{huang-scat}. They
considered the elastic scattering of two-dimensional electrons off a charged
impurity sitting in the middle of a semiconductor quantum well, taking into
account the spin-orbit coupling (SOC) created around the impurity.
They have found that despite the cylindrical symmetry of the electrostatic
potential created by the impurity, the DSCS can be
asymmetric with respect to the forward scattering direction,
and its antisymmetric component is proportional to the out-of-plane
component of the polarization vector of the incoming electron.
This effect is called asymmetric or skew scattering,
and is reminiscent of the so-called Mott skew scattering in three dimensions.
\cite{mott,motz-rmp,engel-review}

This special scattering behaviour caused by the SOC around the impurity
can have a directly measurable consequence on the transport of 
spin-polarized carriers, the so-called anomalous Hall effect (AHE).
The AHE in bulk metals has been in the focus of experimental and theoretical 
research for many decades\cite{karplus-luttinger,smit1,smit2,berger-sidejump}, 
and recent advances in the field of magnetic semiconductors have increased the 
activity within this area even 
further\cite{ohno-prl92,ohno-science98,manyala-natmat,jungwirth-aheinfs,mihaly-ahe}.

Recently Cumings et al\cite{cumings} have observed the AHE in a paramagnetic 
two-dimensional electron gas (2DEG) created in a semiconductor quantum well, 
and its appearance was attributed to the asymmetric electron-impurity scattering. 
In this experiment an out-of-plane magnetic field was applied, resulting in the 
Lorentz force acting on the moving electrons and a finite spin polarization of 
the carriers via the Zeeman effect.
The Lorentz force alone would result in the well-known normal Hall resistivity,
but the simultaneous presence of the finite spin polarization 
and the skew scattering process gives rise to an additional, anomalous Hall 
component.

The experiment of Cumings et al.\cite{cumings} has confirmed that the 
electron-impurity scattering can contribute significantly to the Hall resistivity
in paramagnetic 2DEGs.
Motivated by this fact, in this work we extend the problem considered
by Huang et al.\cite{huang-scat} and study the spin-dependent electron-impurity
scattering process in a 2DEG where the impurity might be located at any 
finite distance from the plane of the 2DEG. 
%The problem under consideration is also strongly related to a recent study\cite{glazov-randomrashba}
%of spin relaxation in a random Rashba field created by impurities in the 
%vicinity of the 2DEG.
%In this work 
We focus our attention to the symmetry properties of the DSCS
describing individual scattering processes and
we show that the properties of the DSCS are fundamentally
different when the impurity is located in the middle of the quantum well and when
it is at a finite distance from that. In the former case skew scattering arises
only if the polarization vector of the incident electron has a finite out-of-plane
component. 
In contrast, we find that in the latter case skew scattering happens provided 
that the polarization vector of the incident electron has a finite projection
on the plane spanned by the normal vector of the plane of the 2DEG and the initial propagation
direction.
In the first part of this paper we summarize the rigorous quantum mechanical 
derivation of this result.
To provide a simple physical picture of this scattering
process we also present a semiclassical analysis of the electron dynamics 
affected by the impurity.
Finally, we discuss a possible experimental setup where the special feature
of the considered scattering process could give rise to a Hall-like
effect in the presence of an \emph{in-plane} magnetic field.

\begin{figure}[t]
\includegraphics[scale=0.4]{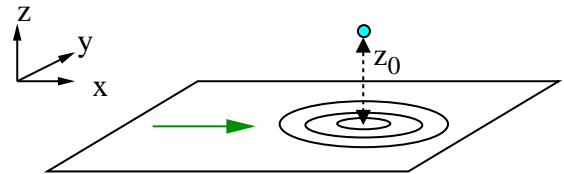}
\caption{\label{fig:2dscat_palyi}
(color online) The electron in a 2DEG
(represented by the green arrow) approaches the impurity located in the position
$\vec r_0 = (0,0,z_0)$. Circular lines represent the equipotentials of the
impurity potential.}
\end{figure}

We consider a 2DEG in the $x$-$y$ plane created in a symmetric quantum well, 
and a charged pointlike impurity in the position $\vec r_0 = (0,0,z_0)$.
The setup is shown in Fig. \ref{fig:2dscat_palyi}. 
Assuming that the SOC strength $\lambda$ is energy-independent and has the same 
value in the quantum well and barrier materials,
the contribution of the impurity to the Hamiltonian is\cite{winkler}
\bnen
\label{eq:himp}
H_{\rm i} = V_{\rm i}(\vec r) + \f{\lambda e}{\hbar} \vec E_{\rm i}(\vec r)
\cdot (\vec \sigma \times \vec p). 
\eden
Here $\vec r$ is three-dimensional coordinate vector of the electron,
$V_{\rm i}$ is the electrostatic potential created by the impurity,
$\vec E_{\rm i} = \vec \nabla V_{\rm i}/e$
is the electrostatic field created by the impurity and 
$\vec \sigma$ is the vector of Pauli matrices representing electron spin.
Note that in the presented theory the screening effects can be 
incorporated into $V_{\rm i}$. 
If screening is taken into account, then $V_{\rm i}$ may loose its spherical 
symmetry around the impurity, however, its cylindrical symmetry around the
$z$ axis is still retained. 

We start with the quantum mechanical analysis of the electron dynamics in this system.
For simplicity the 2DEG is treated as ideal in the sense that electrons are
confined to the plane.
Under this assumption, the system can be modelled by the following 
effective two-dimensional Hamiltonian:
\bean
\nonumber
H_{\rm 2D} &=& \f{p_x^2 + p_y^2}{2m^*} + \bar{V}_{\rm i} + 
\f{\lambda e}{\hbar} \sigma_z \left(
\bar E_{{\rm i},y} p_x - \bar E_{{\rm i},x} p_y
\right) + \\
&+&\f{\lambda e}{2\hbar}\left[
\sigma_x \left\{p_y, \bar E_{{\rm i},z}\right\} - \sigma_y \left\{p_x, \bar E_{{\rm i},z}\right\}
\right], 
\label{eq:h2ddescartes}
\eean
where for any $f \in \{V_{\rm i}, E_{{\rm i},x}, E_{{\rm i},y}, E_{{\rm i},z}\}$
we have defined the notation $\bar{f}(x,y) = f(x,y,0)$,
and $\{.,.\}$ denotes the anticommutator.
This form of the Hamiltonian can be derived in a rigorous way using the standard
dimension reduction technique used by Huang et al\cite{huang-scat}, with the 
negligation of the energy dependence of the effective mass $m^*$ and SOC 
strength $\lambda$.
Note that the cylindrical symmetry of the electrostatic potential $\bar{V}_{\rm i}$ 
and field $\bar{\vec E}_{\rm i}$ created by the impurity is retained even if 
screening effects are incorporated into $V_{\rm i}$.

Having the effective 2D Hamiltonian $H_{\rm 2D}$ in hand, now it is possible
to study the scattering of electrons on the spin-dependent impurity
potential.
In the absence of the impurity, $H_{\rm 2D}$ has plane wave eigenfunctions. 
We will consider the scattering of the electron plane wave 
\bnen
\phi_{\gamma}(\rho,\varphi) = e^{ik\rho\cos\varphi} \gamma,
\eden
which has energy $E = \hbar^2 k^2/2m^*$ and propagates along the $x$ axis.
Here $\rho$ and $\varphi$ denotes the standard planar polar coordinates and
$\gamma$ is a normalized two-component complex vector describing the spin
state of the plane wave. 
We denote the polarization vector of the incident electron by 
$\vec P_0$, which is a three-dimensional real unit vector, and is related to the
spinor $\gamma$ by the expression $\vec P_0 = \gamma^\dag \vec \sigma \gamma$.
Here $\dag$ denotes the combination of complex conjugation and transposition.

It has been shown\cite{palyi-rashbadot} that in two-dimensional 
spin-dependent electron scattering problems
the DSCS can be expressed as the function of the scattering
angle $\varphi$ and the polarization vector of the incident electron $\vec P_0$ in
the following form:
\bnen
\label{eq:sofp}
\sigma_{\rm diff}(\varphi,\vec P_0) = 
c(\varphi) + \vec v(\varphi) \cdot \vec P_0,
\eden
Here the dot represents scalar product, 
and the function $c(\varphi)$ and the vector-valued function
$\vec v(\varphi) = (v_1(\varphi),v_2(\varphi),v_3(\varphi))$ 
is related to $S$-matrix and the scattering amplitude.

It can be shown straightforwardly that the Hamiltonian in Eq. 
\eqref{eq:h2ddescartes} has three important symmetries: $H_{\rm 2D}$
commutes with the out-of-plane component of the total angular momentum operator 
($J_z = -i\hbar \partial_\varphi + \hbar\sigma_z/2$), 
with time reversal ($T = i\sigma_y C$ where $C$ is the complex conjugation) and
with a special combined symmetry of real-space reflection and spin rotation
($\sigma_y P_x$, where $P_x$ is the spatial reflection with respect to the $x$
axis). 
However, if $z_0 = 0$, i.e. the impurity is located in the plane of the quantum well,
then an additional symmetry of the Hamiltonian can be found.
Namely, in this case 
the out-of-plane component $\bar{E}_{{\rm i},z}$ 
of the electric field created by the impurity
vanishes identically, and so does the last term in the Hamiltonian in 
Eq. \eqref{eq:h2ddescartes}.
It means that in this special $z_0 = 0$ case $\sigma_z$ also commutes with
$H_{\rm 2D}$, and as a consequence of that the symmetry properties of the DSCS
 are different in the cases $z_0 = 0$ and $z_0 \neq 0$, as it will be shown
below.

Starting from these symmetry properties of the Hamiltonian $H_{\rm 2D}$ in 
Eq. \eqref{eq:h2ddescartes}, and 
following the method using the $S$-matrix formalism 
outlined in Ref. \onlinecite{palyi-skewpaper} we were able to 
derive the symmetry properties of the functions 
$c$ and $\vec v$ appearing in the formula Eq. \eqref{eq:sofp} of the 
DSCS.
Our findings are summarized in the first and second lines of Table \ref{tab:table}.
In the $z_0 = 0$ case, in correspondence with previous results\cite{huang-scat}, 
we have found that skew scattering can arise only if the out-of-plane component
of the polarization vector of the incident electron is finite.
This can be deduced using the first line of Table \ref{tab:table} and
Eq. \eqref{eq:sofp}.
On the other hand, a fundamentally different behaviour is found in the 
$z_0 \neq 0$ case, when the impurity is lifted out from the 2DEG plane.
In this case, skew scattering is forbidden only if the initial polarization
vector is aligned with the $y$ axis, and the DSCS
can become asymmetric if the inital polarization vector has a finite component
in the $x$-$z$ plane (see the second line of Table \ref{tab:table} and
Eq. \eqref{eq:sofp}). 
In general, in the $z_0 \neq 0$ case skew scattering happens provided that the
initial polarization vector $\vec P_0$ has a finite projection on the plane 
spanned by the normal vector of the 2DEG and the initial propagation direction.

\begin{table}
\begin{ruledtabular}
\begin{tabular}{lcccc}
case & $c$ & $v_1$ & $v_2$ & $v_3$ \\
\hline
exact $z_0 = 0$    & S & 0 & 0 & A\\
exact $z_0 \neq 0$ & S & A & S & A\\
\hline
Born  $z_0 = 0$    & S & 0 & 0 & 0\\
Born  $z_0 \neq 0$ & S & A & S & 0
\end{tabular}
\end{ruledtabular}
\caption{\label{tab:table}Symmetry properties of the quantities determining the
differential scattering cross section in Eq. \eqref{eq:sofp}. 
S (A) denotes the quantities which are even (odd) function of the scattering
angle $\varphi$. 0 denotes the quantities which vanish identically as a result
of the symmetries of the system.}
\end{table}

Usually, the symmetry properties of the DSCS calculated
in the first Born approximation (FBA) are more restrictive than those of the 
exact DSCS. 
A particular example is the $z_0 = 0$ case considered in Ref. \onlinecite{huang-scat},
where it has been shown that the quantity $\vec v(\varphi)$ in Eq. \eqref{eq:sofp} 
characterizing the spin-dependent component of the 
DSCS vanishes if the scattering problem is treated in the FBA.
(Compare first and third lines of Table \ref{tab:table}.)
We have applied a similar analysis in the $z_0 \neq 0$ case and found that in this
case $\vec v(\varphi)$ can be finite in the FBA. 
Moreover, we derived its symmetry properties as well, and 
summarized the results in the fourth line of Table \ref{tab:table}.

%Note that our quantum mechanical results are in
%agreement with the results of the preceding semiclassical analysis.

Now we present a simple semiclassical interpretation of the derived symmetry 
properties characterizing the considered spin-dependent scattering process.
In the followings we use the quantum Hamiltonian in Eq. \eqref{eq:himp}
to derive equations of motion for the observables, and otherwise
we treat $\vec r$, $\vec p$, $\vec \sigma$ as strictly classical
quantities. Specially, instead of $\vec \sigma$ we will use the three-dimensional
unit vector $\vec P$.

Consider a semiclassical particle in the 2DEG approaching the scattering 
centre along the $x$ axis with impact parameter $b$ and spin polarization 
vector $\vec P_0$.
For simplicity we assume that the two-dimensional semiclassical dynamics of the scattered 
electron is determined mainly by the electrostatic potential $V_{\rm i}$, 
and the SOC plays the role of a weak perturbation.
The trajectory of the motion affected only by $V_{\rm i}$ (the `unperturbed' trajectory)
is given by $\vec r(t,b) = (x(t,b),y(t,b),0)$, 
the position vector of the particle with impact parameter $b$ at time $t$.
We assume that at the moment $t =0$ the particle is approaching the scattering centre
but still out of the range of the potential created by the impurity. 
Trivially
\bnsn \label{eq:traj-sym}
 \bean
   x(t,b) &=& x(t,-b),\\
   y(t,b) &=& -y(t,-b),
 \eean
\edsn
i.e. the trajectories corresponding to impact factors 
$b$ and $-b$ are related by a reflection with respect to the $x$ axis.
During the motion the SOC [the second term in Eq. \eqref{eq:himp}]
acts as an effective inhomogeneous magnetic field 
felt by the electron spin: $\vec P \cdot \vec B_{\rm eff}(\vec r,\dot{\vec r})$,
where
\bnen
\label{eq:beff}
\vec B_{\rm eff}(\vec r,\dot{\vec r}) = -\frac{\lambda e m^*}{\hbar}
\left[\vec E(\vec r) \times \dot{\vec r}\right].
\eden
The dot ($\dot{\vec r}$) denotes time derivative, and we used $\dot{\vec r} =  \vec p / m^*$.
With respect to the spin and orbital dynamics, there are two important 
consequences of the presence of this inhomogenous effective
magnetic field.
Firstly, the spin of the moving particle will precess around the effective
magnetic field. 
Secondly, the inhomogeneity of the effective magnetic field gives rise to 
a Stern-Gerlach-like force which deflects the particle from its unperturbed trajectory.

For a given impact factor $b$ and unperturbed trajectory $\vec r(t,b)$, 
the equation of motion for the spin of the moving electron is 
\cite{ballentine}
\bnen
\label{eq:bloch}
\dot{\vec P}(t,b) = \frac{1}{\hbar} \vec P(t,b) \times \vec B_{\rm eff}(\vec r(t,b),\dot{\vec r}(t,b)),
\eden
similarly to the well-known Bloch-equations\cite{kittel}.
If the spin of the incident particle $\vec{P}_0 \equiv \vec{P}(t=0,b)$ is given, and with
this initial value condition the solution of Eq. \eqref{eq:bloch} is known, 
then the Stern-Gerlach force\cite{ballentine}
can be expressed as the gradient of the local SOC energy
\bnen
\label{eq:sgforce}
\vec F(\vec r(t,b),\vec P(t,b)) = - \vec\nabla[\vec P(t,b) \cdot \vec B_{\rm eff}(\vec r(t,b),\dot{\vec r}(t,b))].
\eden
The presence of this force is due to the presence of SOC close to the scattering
centre, and it deflects the particles from their original, unperturbed trajectory.

%\begin{figure}[t]
%\includegraphics[scale=0.4]{skew-sketch.eps}
%\caption{\label{fig:skew-sketch}
%(color online) Semiclassical picture of the skew scattering mechanism. Thin
%lines represent the equipotentials of the potential created by the impurity.
%Black lines show the unperturbed trajectories corresponding to impact factors $b$ 
%and $-b$. The reflection symmetry of these unperturbed trajectories is
%broken by the SOC created by the impurity under certain conditions (see text).
%Green (gray) lines show the trajectories affected by the SOC.}
%\end{figure}

We claim that if the spin of the incident electron has a finite projection on 
the $x$-$z$ plane then the presence of the SOC destroys the reflection symmetry 
of the motion with respect to the $x$ axis and therefore gives rise to an 
asymmetry in the DSCS as well.
This is a consequence of the fact that if the polarization vector lies
in the $x$-$z$ plane then the components of the Stern-Gerlach force 
fulfill the relations
\bnsn \label{eq:classth}
 \bean
  F_1(\vec r(t,b),\dot{\vec r}(t,b)) &=& -F_1(\vec r(t,-b),\dot{\vec r}(t,-b)),\\
  F_2(\vec r(t,b),\dot{\vec r}(t,b)) &=&  F_2(\vec r(t,-b),\dot{\vec r}(t,-b)).
 \eean
\edsn
(Here and henceforth the vector subscripts 1, 2 and 3 are equivalent to 
$x$, $y$ and $z$, respectively.)
We sketch the steps of the proof in the followings.
The definition of $\vec B_{\rm eff}$ in Eq. \eqref{eq:beff}, the properties in
Eqs. \eqref{eq:traj-sym} and the cylindrical symmetry of $\vec E(\vec r)$ implies
symmetry relations of the components of $\vec B_{\rm eff}$:
\bnen
\label{eq:beffsym}
[\vec B_{\rm eff}(\vec r(t,b),\dot{\vec r}(t,b))]_k = 
(-1)^k [\vec B_{\rm eff}(\vec r(t,-b),\dot{\vec r}(t,-b))]_k,
\eden
where $k = 1,2,3$.
The Picard-Lindelof\cite{apostol} solution of 
Eq. \eqref{eq:bloch} is
\bean
\label{eq:plsol}
\vec P(t,b) &=& \vec P_0 +
\int_0^t dt'    \frac{\vec B_{\rm eff}(t'  ,b)}{-\hbar} \times \vec P_0 +\\ \nonumber
&&\int_0^t dt' 
\int_0^{t'}dt'' \frac{\vec B_{\rm eff}(t' ,b)}{-\hbar} \times
                \frac{\vec B_{\rm eff}(t'',b)}{-\hbar} \times \vec P_0 + \dots,
\eean
where we used the notation 
$\vec B_{\rm eff}(t,b) \equiv \vec B_{\rm eff}(\vec r(t,b),\dot{\vec r}(t,b))$
for brevity.
Using Eqs. \eqref{eq:beffsym} and \eqref{eq:plsol}, and assuming
an initial spin $\vec P_0 \perp (0,1,0)$, it is straighforward to prove that
\bnen
\label{eq:sigmasym}
P_k(t,b) = -(-1)^{k}P_k(t,-b).
\eden
Finally, substituting Eqs. \eqref{eq:beffsym} and \eqref{eq:sigmasym} into Eq. \eqref{eq:sgforce}
results in Eqs. \eqref{eq:classth}.
A similar analysis shows that if the spin of the incident electron is 
aligned with the $y$ axis
then the reflection symmetry between the trajectories corresponding
to $b$ and $-b$ is retained in the presence of SOC, and therefore
this is the only case when skew scattering does not take place.
As a generalization of these results, we formulate the central theorem of this
semiclassical analysis as follows: if the initial polarization
vector has a finite projection on the plane spanned by the normal of the 2DEG 
and the initial propagation direction, then the DSCS becomes \emph{asymmetric} with
respect to forward scattering direction.
Note that this conclusion is equivalent to the results of the rigorous
quantum mechanical symmetry analysis presented above.

By repeating the preceding analysis for the case when the impurity
is located in the plane of the 2DEG ($z_0 = 0$), and using the fact that
in this case the out-of-plane component of the electric field vanishes identically,
it can be shown that skew scattering may occur only if the out-of-plane component
of the initial polarization vector of the electron is finite.
This result is in agreement with the quantum mechanical result
summarized in the first line of Table \ref{tab:table}.

\begin{figure}[t]
\includegraphics[scale=0.4]{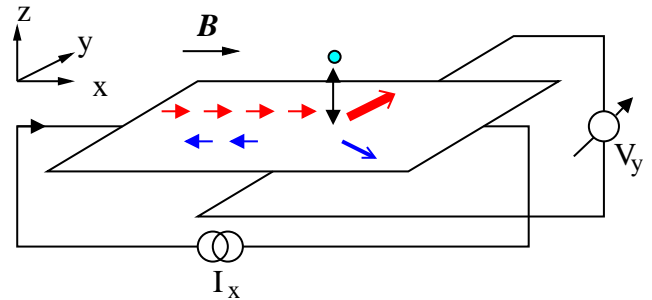}
\caption{\label{fig:2dnkahe}
(color online) Proposed measurement setup for experimental investigation of the
Hall-like effect if an impurity plane is present in the sample at a finite 
distance from the 2DEG plane.
The short horizontal red (blue) 
arrows in the figure represent spins parallel (antiparallel)
to the polarizing magnetic field $\vec B$.}
\end{figure}

So far we have discussed the symmetry properties of the DSCS corresponding
to individual scattering events. 
Here we argue that in specially prepared disordered samples this peculiar
scattering mechanism may give rise to an experimentally observable effect
similar to the skew scattering induced AHE in spin-polarized systems.
The setup is shown in Fig. \ref{fig:2dnkahe}.
The sample for the proposed experiment should contain a symmetric quantum
well with an additional delta-doped impurity layer at a finite distance from the 
quantum well.
For clarity in Fig. \ref{fig:2dnkahe} we show only a single impurity.
We also assume that the 2DEG is fully or partially spin-polarized by a static 
homogeneous in-plane magnetic field $\vec B$.
If a finite dc current parallel to the magnetic field is flowing through this 
sample, then the skew scattering mechanism will result in a finite Hall signal
despite the fact that the magnetic field has no out-of-plane component:
skew scattering means that the electrons drifting in the direction 
of the driving electric field 
are preferably scattered by impurites to, say, the left (right) if their spin is
parallel (antiparallel) to the magnetic field.
Therefore in the polarized 2DEG more electrons will pile up at the left
edge of the sample than at the other edge, giving rise to a finite transversal
bias between the two edges.
Though quantitative estimates regarding the magnitude of this effect are not
presented here, the fact that $v_1$ is finite in the FBA while $v_3$ vanishes
suggests that the predicted effect should be at least comparable with
the anomalous Hall effect\cite{cumings} measured in the presence of an
out-of-plane polarization.

In summary, we have shown that the spin-dependent electron-impurity
scattering in a 2DEG has different symmetry properties when the impurity
is sitting in the plane of the 2DEG ($z_0 = 0$) and when it is located at 
a finite distance from that ($z_0 \neq 0)$. 
In the former case skew scattering can arise only if the spin of the incident
electron $\vec P_0$ has a finite out-of-plane component, whereas in the latter
case skew scattering happens if $\vec P_0$ 
has a finite projection on the plane spanned by the normal vector of the
2DEG and the initial propagation direction.
%We have also shown that the spin-dependence of the differential scattering 
%cross section appears in the first Born approximation if $z_0 \neq 0$, in contrast
%to the $z_0 = 0$ case. 
We also proposed a measurement setup which could be used to 
experimentally investigate the Hall-like effect arising as a consequence of 
the peculiar scattering mechanism studied in this paper and the presence of
an in-plane magnetic field.

J. Cs. acknowledges the support of the Hungarian Science Foundation OTKA under
the contracts No. T48782 and 75529.

%===

%\bibliography{amott-letter}

\end{document}